# Uncovering Thermal and Electrical Properties of Sb₂Te₃/GeTe Superlattice Films

Heungdong Kwon[1], Asir Intisar Khan[2], Christopher Perez[1], Mehdi Asheghi[1], Eric Pop[2,3], Kenneth E. Goodson[1,3,*]

[1]Dept. of Mechanical Eng., Stanford University, Stanford, CA 94305, U.S.A.
[2]Dept. of Electrical Eng., Stanford University, Stanford, CA 94305, U.S.A.
[3]Dept. of Materials Science & Eng., Stanford University, Stanford, CA 94305, U.S.A.

*E-mail: goodson@stanford.edu

**ABSTRACT: Superlattice-like phase change memory (SL-PCM) promises lower switching current than conventional PCM based on Ge₂Sb₂Te₅ (GST). However, a fundamental understanding of SL-PCM requires detailed characterization of the interfaces within such a SL. Here, we explore the electrical and thermal transport of SLs with deposited Sb₂Te₃ and GeTe alternating layers of various thicknesses. We find up to ~4× reduction of the effective cross-plane thermal conductivity of the SL stack (as-deposited polycrystalline) compared to polycrystalline GST (as-deposited amorphous and later annealed) due to the thermal interface resistances within the SL. Thermal measurements with varying periods of our SLs show a signature of phonon coherence with a transition from wave-like to particle-like phonon transport, further described by our modeling. Electrical resistivity measurements of such SLs reveal strong anisotropy (~2000×) between the in-plane and cross-plane directions due to the weakly interacting van der Waals gaps. This work uncovers electro-thermal transport in SLs based on Sb₂Te₃ and GeTe, for improved design of low-power PCM.**

In recent years, there has been a large and growing interest in new materials for new data storage opportunities. These opportunities are driven in equal parts by growing memory demands of artificial intelligence applications, and by shortcomings of existing memory technologies such as flash (non-volatile, but relatively slow) and DRAM (fast, but volatile).[1,2] In this context, phase change memory (PCM) has become a key emerging technology,[3] with Intel and Micron jointly introducing a commercial PCM-based product called 3D X-point in 2017. However, PCM improvements remain necessary, including reduction of resistance drift, higher endurance, and lower switching current density, ideally < 1 MA cm⁻².

Several recent studies have suggested that replacing the phase-change material (conventionally Ge₂Sb₂Te₅, i.e. GST) with a superlattice (SL) of alternating thin chalcogenide films (e.g., Sb₂Te₃ and GeTe) could offer lower switching power, faster switching speed, and higher endurance.[4,5] Interestingly, the fundamental operation of SL-PCM is not well understood. It was initially proposed that its memory states are controlled at the internal interfaces by subtle changes in the position of Ge atoms in multi-layered Sb₂Te₃/GeTe films.[6] However, the broader influence of internal interfaces on electrical and thermal transport in SL-PCM remains poorly understood, especially when considering that conventional PCM is driven by a thermally-induced phase change mechanism. In addition, the complex physics of energy and charge transport at these interfaces also presents an opportunity to better understand the mechanism of SL-PCM device switching.

In the past, both our groups and others have demonstrated the benefits of thermal engineering of PCMs (e.g. by increasing their thermal resistance) by introducing multi-layer dielectric insulation,[7-9] graphene,[10] $MoS_2$,[11] and other interface layers[12,13] to reduce the switching current and power density. The influence of PCM interfaces and of ballistic heat flow[14] are only expected to become magnified in ultrathin nanoscale films and devices, and particularly in SL structures which have numerous *internal* interfaces.[15] In fact, recent results suggest that the low switching current in SL-PCM could be thermally driven due to the thermal resistance of interfaces within the PCM stack.[14,16]

Motivated by the need for a deeper understanding of SL-PCM materials, here we report the thermal and electrical transport characterization of $Sb_2Te_3$/GeTe SLs by measuring their temperature- and period-dependent cross-plane thermal conductivity. These measurements uncover a minimum thermal conductivity and a transition from coherent, wave-like phonon transport to quasi-ballistic phonon transport around the 3/1 nm/nm period. Thickness-dependent measurements also reveal the thermal boundary resistance between SL-PCM and TiN, a typical PCM electrode material. Finally, we report electrical resistivity measurements of 4/1 nm/nm $Sb_2Te_3$/GeTe SLs, revealing a strong, nearly 2000-fold anisotropy between the cross-plane and in-plane directions.

The cross-plane thermal measurements are carried out using time-domain thermoreflectance (TDTR),[17] which is shown with the prototypical SL stack in Figure 1a. TDTR uses a pump laser which repeatedly heats the top metal surface (here Pt) and a probe laser which measures its temperature. Modulating the pump laser beam from 1 to 10 MHz allows changing its penetration depth into the SL stack and the Si substrate. The measured thermal decay trace is fit to the exact solution of the radial-symmetric heat diffusion equation of the system, enabling the extraction of thermal conductivity and interface thermal resistances within the SL, as well as the thermal boundary resistance of the SL interface with TiN. (Additional details are in Supporting Information Section 1 and in Figure S1.) In addition, we equip our TDTR system with a high vacuum heating stage to enable temperature-dependent measurements.

The $Sb_2Te_3$/GeTe SLs used in this work are carefully prepared using RF magnetron sputtering, followed by transmission electron microscopy (TEM), as shown in Figure 1b. As a pre-treatment for the (poly)crystalline $Sb_2Te_3$/GeTe deposition, a 10-minute Ar plasma cleaning process is performed on a silicon wafer with high resistivity, followed by 4 nm $Sb_2Te_3$ seed layer deposition at room temperature (RT). This *in situ* cleaning process is used to remove carbon residues and any native oxide on the substrate surface, and is also essential for depositing PCM films on the bottom electrode of PCM devices to produce better interfaces.[12] The temperature in the chamber is then elevated to 180 °C without breaking vacuum and the alternating multilayers of GeTe and $Sb_2Te_3$ are deposited at a pressure of 4 mTorr.

The cross-sectional high-angle annular dark field (HAADF) scanning transmission electron microscope (STEM) images of the resulting SL (Figure 1b) reveal sharp interfaces with clear van der Waals gaps. Although alternating thin layers of $Sb_2Te_3$ and GeTe were deposited, the STEM images show a reconfiguration of the superlattice into 5-atom layers of $Sb_2Te_3$ and odd-numbered (7-, 9-, or 11-atom) layers of $Ge_xSb_yTe_z$ during the sputter deposition at elevated temperature.[18,19,20] However, nearly perfect alignment of all vdW gaps parallel to the substrate is maintained, as evident from the STEM. We also note the presence of (some) stacking faults and other interfacial defects.[21] These unique structural attributes are expected to play a role in cross-plane electrical and thermal transport, as described below. We cap the SL stacks *in situ* with 10 nm of TiN, the same material used for PCM electrodes, followed by a layer of 70 nm Pt, which is used for the TDTR measurement.

Figure 1c displays our measured effective cross-plane thermal conductivity of a ~60 nm SL with 12 deposited periods of 4 nm $Sb_2Te_3$ and 1 nm GeTe, with a 10 nm TiN capping layer, from RT to 400 °C. This effective thermal conductivity includes the contribution from the thermal resistances of the SL layers as well as the thermal boundary resistance at the SL/TiN and SL/Si interface. The SL is polycrystalline as-deposited at 180 °C and the overall SL stack has larger effective thermal conductivity ($0.29 \pm 0.02$ W m$^{-1}$ K$^{-1}$) at RT. As a baseline, we also depict the thermal conductivity of as-deposited amorphous 60 nm thick $Ge_2Sb_2Te_5$ (a-GST, including a similar capping layer) ~0.17 W m$^{-1}$ K$^{-1}$ at RT.[22] The thermal conductivity of GST sharply increases at ~150 °C during the transition to its face-centered cubic phase and continues to increase up to ~1.05 W m$^{-1}$ K$^{-1}$ at the onset of transition to its hexagonal crystalline phase at 200 °C. [22] The thermal conductivity of hexagonal-GST slightly increases to ~1.24 W m$^{-1}$ K$^{-1}$ at 350 °C. The reported experimental data for 350 nm thick hexagonal-GST film indicates that the thermal conductivity decreases with decreasing temperature from ~2.2 W m$^{-1}$ K$^{-1}$ at ~350 °C to ~1.7 W m$^{-1}$ K$^{-1}$ at RT.[22] The linear but gradual increase in the thermal conductivity of the hexagonal-GST from RT to 350 °C is attributed to the contribution of electrons to the energy transport in the hexagonal-GST films.[22,23] This weak temperature dependency is also observed in our SL film where the effective thermal conductivity of the SL increases only slightly with temperature, from 0.29 W m$^{-1}$ K$^{-1}$ at RT to a maximum of $0.47 \pm 0.05$ W m$^{-1}$ K$^{-1}$ at ~400 °C. The effective cross-plane thermal conductivity of the SL stack is 5× lower than the *crystalline* GST phases at high temperature, which is primarily attributed to the strong role of the interfaces and vdW gaps within the SL. In addition, the as-deposited polycrystalline SL films may have smaller grain size than the annealed GST (as-deposited amorphous), which could also contribute to the measured thermal conductivity difference above.

Next, we wish to quantify the thermal conductivity of the $Sb_2Te_3$/GeTe SL itself, and the thermal boundary resistance (TBR) between the SL and the TiN electrode, $R_{SL/TiN}$. These thermal resistances can have a large impact on the temperature rise within SL-PCMs and may in part shed light on the underlying switching mechanism in these memory devices. For this purpose, we design a set of ~14 nm, 29 nm, 59 nm, and 89 nm thick samples comprised of repeating 1/4 nm/nm GeTe/Sb$_2$Te$_3$ layers on top of the 4 nm Sb$_2$Te$_3$ seed layer. The top TiN/Pt capping and bottom 4 nm Sb$_2$Te$_3$ seed layers on the silicon wafer are identical for all samples. The total thermal resistances of these SL films, $R_{tot}$, is measured using TDTR at RT and plotted vs. the film thickness in Figure 1d. The slope of the linear regression for $R_{tot}$ can be used to estimate the intrinsic cross-plane thermal conductivity of the SL, $k_{SL} \approx$ 0.41 ± 0.04 W m$^{-1}$ K$^{-1}$, at room temperature.

On the other hand, the vertical intercept as the number of film interfaces approaches zero yields the TBR between the TiN capping layer and SL-PCM film including a smaller contribution from the SL/Si interface, $R_{SL/TiN+SL/Si} \approx$ 52.4 m$^2$ K GW$^{-1}$, because the TDTR signal is more sensitive to the top interface with TiN rather than that the bottom interface with the Si substrate. To confirm this, we conducted the exact fitting process to separately estimate $R_{SL/TiN}$ = 41.7 ± 10.8 m$^2$ K GW$^{-1}$ (see Supporting Information Section 2, Figure S2, and Table S1). This result is ~47% larger than what has been reported between TiN and fcc-GST $R_{GST/TiN}$ ~28 m$^2$ K GW$^{-1}$.[24] More detailed interface characterization would be needed to explain this difference, but this could be partially attributed to the different deposition temperature of our SL-PCM films (~180 ℃), while the fcc-crystalline interface could form a rougher interface compared to an amorphous Ge$_2$Sb$_2$Te$_5$ film. In the Supporting Information Section 3, we have also detailed a theoretical analysis that indicates the acoustic velocity of the SL film could be suppressed by ~10% due to the creation of minibands and a flattened phonon dispersion spectrum in SLs.[25] This alone could account for about 27% larger thermal resistance between the SL film and TiN, compared to GST and TiN.

To assist with a better understanding of the switching mechanisms and complex physics of energy transport at the interfaces of SL-PCM, we have conducted systematic thermal characterization of ~60 nm thick films with different size periods and thickness ratios (t.r.) of Sb$_2$Te$_3$/GeTe (Figure 2). All the films were capped *in situ* with 10 nm thick TiN layers. The SL-PCM period sizes of Sb$_2$Te$_3$/GeTe are 24/6, 16/4, 12/3, 8/2, 4/1, and 2/0.5 nm/nm (all with t.r. of 4:1), 3/1 nm/nm (t.r. 3:1), 2/1 nm/nm (t.r. 2:1), and 1/1 nm/nm (t.r. 1:1). As shown in Figure 2a, the cross-plane thermal conductivity of SL-PCM films decreases from 0.69 ± 0.06 W m$^{-1}$ K$^{-1}$ for a 24/6 nm/nm period (4 interfaces) to 0.37 ± 0.03 W m$^{-1}$ K$^{-1}$ for a 3/1 nm/nm period (30 interfaces). We note that the thermal conductivity of the SLs shown in Figure 2a excludes the contribution of the TiN capping layer and Si substrate (see Supporting Information Section 2), and thus exclusively represents the thermal transport in SL-PCM films. It is notable that the theoretical minimum lattice thermal conductivity for these materials,[26] $k_{min} \approx$ 0.31 W m$^{-1}$ K$^{-1}$ (see Supporting Information Section 4) is only ~18% smaller than the minimum thermal conductivity of the 3/1 nm/nm period (30 interfaces) SL film. Interestingly, the measured thermal conductivity of our SLs increases again for shorter periods of 3/1 nm/nm, 2/1 nm/nm, and 1/1 nm/nm, corresponding to 40, 48 and 60 interfaces, respectively. This counterintuitive trend suggests a transition of phonon transport behavior from the quasi-ballistic regime (> 4 nm periods) to a wave-like, quasi-coherent regime (2, 2.5, and 3 nm periods), a phenomenon usually observed in high-quality SL films.[27]

To model this observed thermal conductivity behavior of our $Sb_2Te_3$/GeTe SLs, we use the Simkin-Mahan Model (SMM)[28] with a correction for imperfections at the interfaces[29] as detailed below (also see Supporting Information Section 5). The main input parameters to the SMM are the mass ratio of lattice units, the number of lattices consisting of the one period of $Sb_2Te_3$/GeTe, and the bulk phonon mean free path scaled by the lattice size. We first estimate the bulk phonon mean free path $\lambda$ in the SL from conventional kinetic theory $\lambda \approx 3k/(C_v v)$, where $k$ is the phonon thermal conductivity obtained by averaging the bulk thermal conductivity of $Sb_2Te_3$ and GeTe (with 4:1 thickness ratio) in the cross-plane direction, $C_v$ is the volumetric specific heat, and $v$ is the average sound velocity (See Supporting Information Section 3 for detailed calculation), yielding $\lambda \approx 1$ nm. However, this mean free path is shorter than our considered SL periods, and thus the kinetic theory estimate cannot be applied to understand the wave-like phonon behavior shown in Figure 2a.

To go a step further, we recall that layered materials are strongly anisotropic, which could lead to a significant underestimation of the phonon mean free path by kinetic theory,[15,30] due to the change in the phonon density of states in the different directions of the lattice. In addition, time-resolved measurements have suggested that phonons could have relatively longer lifetimes in $Sb_2Te_3$/GeTe SL films.[31] Therefore, to model the experimentally observed thermal conductivity trend (Figure 2a), we use the "modified" SMM[29] (M-SMM) in Figure 2b by employing two free parameters, the bulk phonon mean free path of the SL and the TBR originating from imperfections at SL interfaces. Using the M-SMM, we first determine the best fit (blue solid line) to our experimental data (red filled diamonds representing 24/6, 16/4, 12/3, 8/2, 4/1, and 2/0.5 nm/nm; t.r. 4:1 for $Sb_2Te_3$/GeTe) with $\lambda = 9.4$ nm and the thermal boundary resistance per interface due to the inelastic scattering caused by imperfections alone, $R_{imp.} = 2$ $m^2$ K $GW^{-1}$. Thus, the M-SMM with a longer mean free path compared to kinetic theory estimates better predicts the behavior of the thermal conductivity of our SL-PCM films. This suggests that the mean free path along the cross-plane (c-axis) is larger than the one obtained from the simple kinetic theory, which has also been observed before in similar layered materials such as $MoS_2$ and graphite.[15,30]

Next, we model the thermal conductivities with a fixed $\lambda = 9.4$ nm while varying $R_{imp}$ from 0 to 3 $m^2$ K $GW^{-1}$. We note that for the fits in Figure 2b we only use the SLs with t.r. 4:1 of $Sb_2Te_3$/GeTe, where the data for the SL with largest size period 24/6 nm/nm is available; this is to minimize the impact of the TBR between SL interfaces. The fit for $R_{imp} = 0$ (using SMM) in Figure 2b represents perfect internal interfaces of $Sb_2Te_3$/GeTe and captures the trend in the experimental thermal conductivity data of SLs in Figure 2a. However, this fit clearly overestimates the thermal conductivity for smaller SL periods. Considering $R_{imp} = 1$-$2$ $m^2$ K $GW^{-1}$, our model closely captures the experimental thermal conductivity data. Such low $R_{imp.}$ values indicate overall high-quality interfaces of our SLs (further detailed in Supporting Information Section 5 and in Figure S3), with any contributions most likely arising from the stacking faults[18,32] which are apparent in Figure 1b. We further note that the fitted $R_{imp}$ values agree well with the molecular dynamics simulation results[33] for the thermal boundary resistances per interface (1 to 2 m$^2$ K GW$^{-1}$) originating from imperfections at the three different atomic layers (Te$_1$, Bi, and Te$_2$) of Bi$_2$Te$_3$.

We now turn to electrical resistivity characterization of our SL films, which may also impact their programming in SL-PCM data storage devices. For the in-plane electrical resistivity measurements of thin films, the standard transfer length method (TLM) is usually employed,[34] as shown in Figure 3a. This measurement is sensitive to the in-plane electrical resistivity, $\rho_{\parallel}$, and the electrical contact resistivity with the metal pads, $\rho_C$. For cross-plane electrical resistivity measurements, we utilize a modified TLM (M-TLM), which uses mesa structures to determine the cross-plane electrical resistivity simultaneously.[35] To fabricate the M-TLM structure, we etch into the SL film between the two metal contacts (here Pt) as shown in Figure 3b (also see Supporting Information Section 6). This will increase the contribution of the cross-plane electrical resistance, $R_{e\perp}$, that is proportional to the etched depth, $t_{etch}$.

The contact and cross-plane resistivities of the SL film can be estimated from simultaneously fitting the measured electrical resistance as a function of contact spacing and etch depth, in the modified TLM devices. Figure 3c shows the fitting line of the electrical resistance in each M-TLM device with a SL formed by 4 nm Sb$_2$Te$_3$ and 1 nm GeTe deposition, and Figure 3d shows the total area-specific resistance, $\rho_{etch}$ (including the SL/metal contact resistance), corresponding to the amount of the etched depth. The slope in Figure 3d represents the cross-plane electrical resistivity of these SLs, $\rho_\perp \approx 1.1 \pm 0.35$ $\Omega$ cm at room temperature. The in-plane electrical resistivity is estimated as $\rho_{\parallel} \approx 5.8 \pm 0.3 \times 10^{-4}$ $\Omega$ cm from the lateral TLM. The error bars of $\rho_{\parallel}$ and $\rho_\perp$ are calculated using a Monte Carlo approach,[36] propagated from the uncertainties in the metal contact width and spacing ($\pm 0.5$ μm), as well as the etch depth ($\pm 2$ nm). We note that the measured electrical resistances for different etched depths of SL films show linearity (Figure 3c) across large distances of neighboring contact spacing (up to 20 μm), indicating good uniformity of our sputtered SLs over a large area.

Our measured in-plane electrical resistivity is similar to that previously reported in-plane for such SLs[37] and to crystalline GeTe,[38] but smaller than crystalline Sb$_2$Te$_3$,[39] and nearly 2000-fold smaller than the SL cross-plane resistivity. This strong anisotropy in the electrical transport is in close agreement with previously reported data[40] for such SL films. The large anisotropy is not surprising in these layered SL materials (see Supplementary Table S2), with covalent bonds in-plane and numerous weak van der Waals bonds and some stacking disorder in the cross-plane direction.[40-42] Some charge accumulation or depletion in adjacent SL layers (with smaller and larger band gaps, respectively)[43] could also contribute to the anisotropy of electrical resistivity in these films, and could be the subject of future exploration.

We also note how these SL resistivity values compare to those in a-GST ($> 10^2$ $\Omega$ cm), fcc-GST (~0.1 $\Omega$ cm), and hexagonal-GST (~$10^{-3}$ $\Omega$ cm) near room temperature (RT).[44] In other words, the cross-plane resistivity of the SL is larger than for the crystalline phases of GST while the in-plane resistivity of the SL is slightly smaller than hexagonal-GST. The in-plane SL resistivity can also be used to estimate the *electronic* contribution to the in-plane thermal conductivity with the Wiedemann-Franz law, $k_{e,\parallel} \approx 1.24$ W m$^{-1}$ K$^{-1}$, which represents a lower bound. The total in-plane thermal conductivity of the SL could be up to ~5 W m$^{-1}$ K$^{-1}$ at RT including the phonon component, based on the in-plane thermal conductivity of both crystalline GeTe[38] and Sb$_2$Te$_3$[39] (but likely to be somewhat smaller for the SL due to finite size effects in our ultrathin films). This means that our SLs with 4/1 nm/nm periods have greater than 4-fold (and possibly up to 16-fold) thermal anisotropy between the in-plane and cross-plane directions.

In summary, we reported the electrical and thermal transport characterization of SL-PCMs with Sb$_2$Te$_3$/GeTe alternating layers. The measured in-plane and cross-plane electrical resistivities of the SLs indicate a very anisotropic structure, which can largely impact the electric field and resulting heat generation in SL-based phase change memory (PCM) devices. We also observe up to ~4× lower effective cross-plane thermal conductivity of (poly)crystalline Sb$_2$Te$_3$/GeTe SLs compared to crystalline GST, due to the presence of SL interfaces. These could play a strong role in the reduction of SL-PCM programing current compared to conventional PCM based on GST. We also observed a transition from quasi-coherent to quasi-ballistic phonon transport as we varied the period of Sb$_2$Te$_3$/GeTe films (from 1/1 to 24/6 nm/nm), with a minimum thermal conductivity observed around the 3/1 nm/nm period. Future work could provide additional in-plane and cross-plane thermal and electrical measurements of SLs at cryogenic and high temperatures, to further our understanding of transport physics in such materials.

**ASSOCIATED CONTENT**

**Supporting Information**

The Supporting Information is available free of charge on the ACS Publications website at DOI: XXXX

Additional details on the time domain thermoreflectance measurement method, sensitivity analysis, theoretical calculation of thermal boundary resistance, theoretical estimation of minimum thermal conductivity, additional details on the thermal modeling, additional details and benchmarking of the electrical resistivity measurement

**ACKNOWLEDGMENTS**

We acknowledge the financial support from the Semiconductor Research Corporation, SRC (task 2826.001), Kwanjeong Educational Foundation Fellowship, and Stanford Graduate Fellowship. H.K and A.I.K are thankful to Dr. James McVittie for lab assistance and useful discussion. Authors are grateful to Kazuhiko Yamamoto, Shosuke Fujii and Kazunari Ishimaru from Kioxia Corporation, Japan for obtaining the high-quality scanning transmission electron microscopy images and management support. Part of this work was performed at the Stanford Nanofabrication Facility (SNF) and Stanford Nanofabrication Shared Facilities (SNSF), supported by the National Science Foundation under award ECCS-2026822.

**AUTHOR CONTRIBUTION**

H. K. and A.I.K. conducted the thermal and electrical measurements. A.I.K. and H.K. performed the film depositions. All the authors contributed to the discussion of the experimental results and the writing of the manuscript.

**COMPETING INTERESTS**

The authors declare no competing interests.

**TABLE OF CONTENTS (ToC) FIGURE**

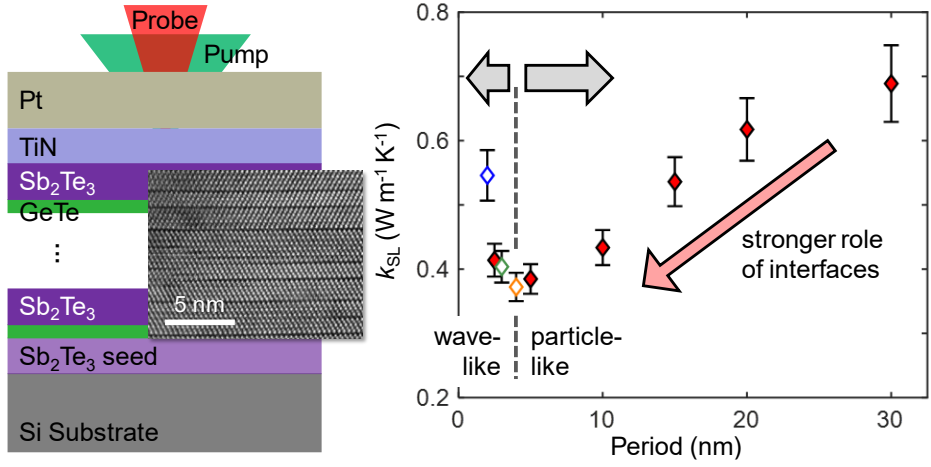

**Figures**

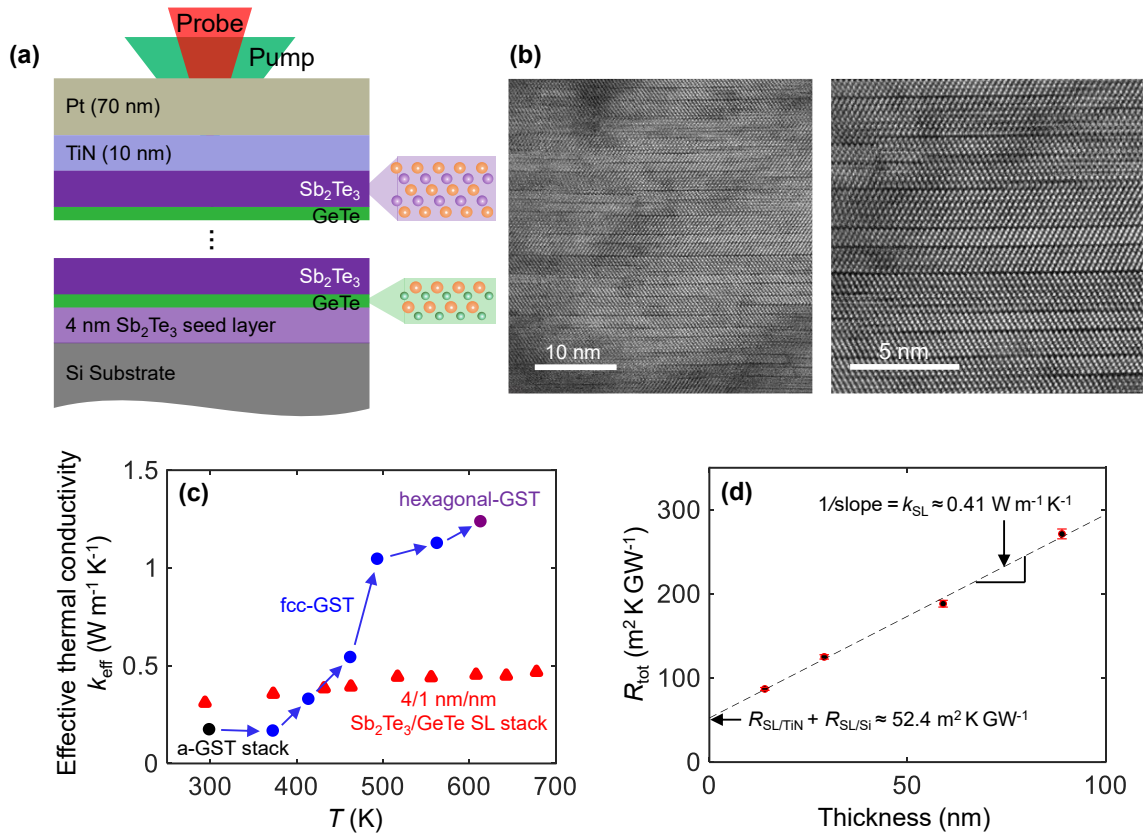

**Figure 1.** Cross-plane thermal characterization of Sb$_2$Te$_3$/GeTe superlattices (SLs). (a) Measurement setup showing, from bottom to top: silicon (Si) substrate, 4 nm Sb$_2$Te$_3$ seed layer, deposited alternating GeTe and Sb$_2$Te$_3$ layers, TiN/Pt capping layers, and pump/probe lasers. (b) High-resolution scanning transmission electron microscopy (STEM) cross-sections of resulting superlattice on a 100 mm diameter Si wafer, revealing sharp interfaces and clear van der Waals gaps. The resulting superlattice is somewhat different from the as-deposited layers, due to interfacial reconstruction which is known to occur during deposition kinetics.[18-20] (c) Measured effective thermal conductivity of 60 nm thick as-deposited SL including the TiN capping layer (red filled triangles) from room temperature to 400 ºC, and the temperature-dependent effective thermal conductivity of Ge$_2$Sb$_2$Te$_5$ (GST) (including similar capping layer) from Ref. 22 (filled circles). The measurement laser spot size is ~10 μm, and the SL measurements are averaged from three locations on the same sample. (d) Measured total thermal resistance of 4/1 nm/nm superlattice with increasing number of periods, from 2 to 17 at room temperature (2 periods = 4 internal interfaces including the bottom Sb$_2$Te$_3$ seed layer = 14 nm thick, etc.). The inverse of the slope of the linear fit estimates the intrinsic thermal conductivity of the SL, and the y-intercept yields the sum of thermal boundary resistance between SL/TiN at the top and SL/Si at the bottom.

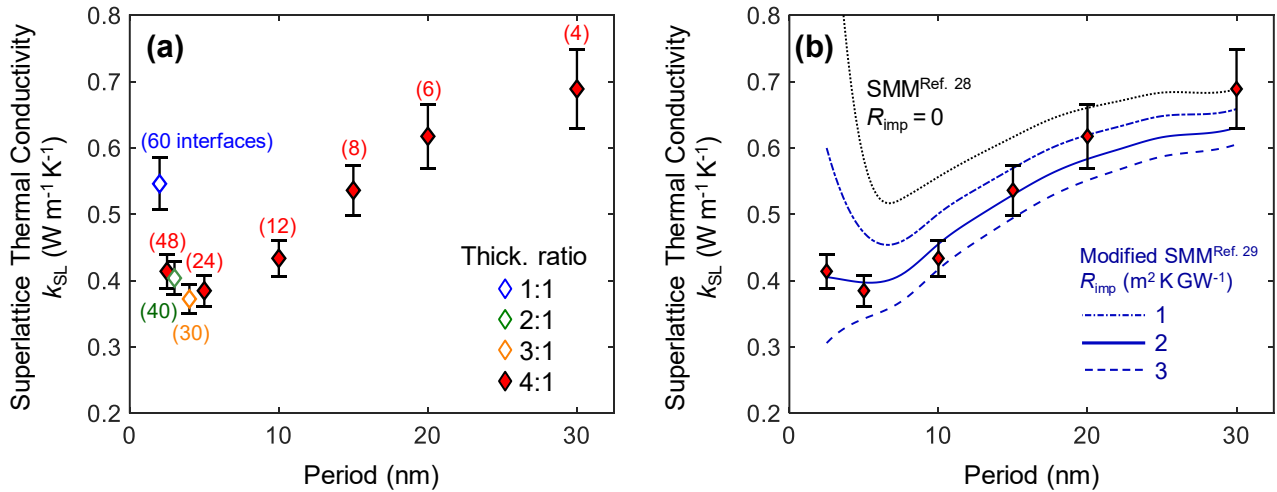

**Figure 2.** (a) Measured cross-plane thermal conductivity of 60 nm thick $Sb_2Te_3$/GeTe superlattice (SL) films as a function of period size ranging from 1/1 to 24/6 nm/nm and of varying thickness ratios, at room temperature. Red filled diamonds represent samples with a 4:1 thickness ratio (i.e., $Sb_2Te_3$/GeTe 4/1, 8/2 nm/nm, etc.). Empty blue, green, and orange diamonds correspond to 1/1, 2/1, and 3/1 nm/nm periods, respectively. The numbers in parentheses indicate the total number of internal interfaces in the respective SL-PCM stack. (b) Data for measured cross-plane thermal conductivity of SL-PCM with a 4:1 thickness ratio of $Sb_2Te_3$/GeTe (red diamonds) and fits using the Simkin and Mahan model (SMM)[28] (dotted black) and Modified SMM[29] (in dotted blue lines). The fits are obtained using a fixed phonon mean free path of ~9.4 nm while the contribution of imperfections to thermal boundary resistance per interface, $R_{imp}$ is varied from 0 to 3 $m^2\,K\,GW^{-1}$. The best fit curve to the experimental data corresponds to $R_{imp} = 2\,m^2\,K\,GW^{-1}$ with a bulk mean free path of 9.4 nm.

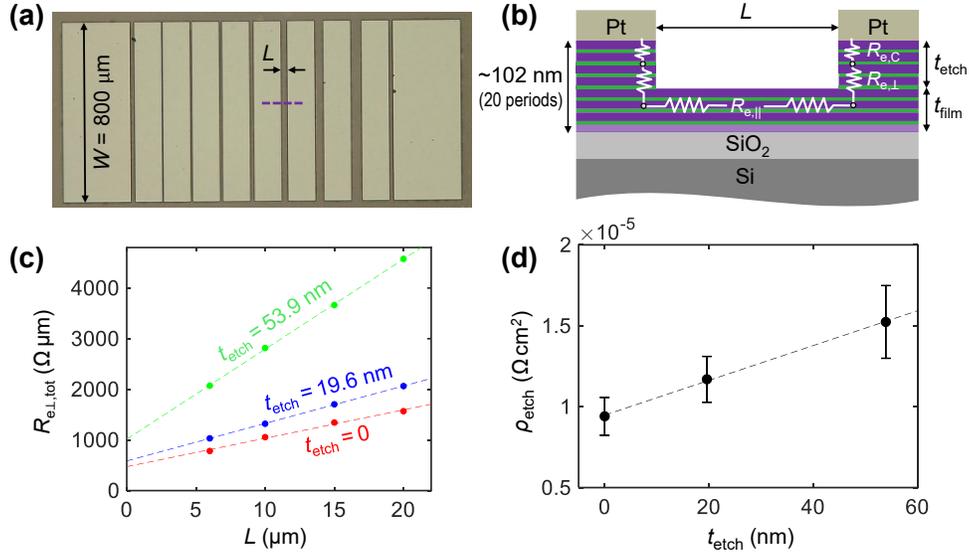

**Figure 3.** Electrical characterization of $Sb_2Te_3$/GeTe superlattices with 4/1 nm/nm period. (a) Top view optical image of transfer length method (TLM) structure, where $L$ is the distance between the Pt contact pads and $W$ is the width of each contact pad. (b) Schematic cross-section of the modified TLM (M-TLM) along the dashed line in (a), with ~82 nm thick Pt contact pads and a 90 nm thick $Sb_2Te_3$/GeTe film with a 4/1 nm/nm period on thermally grown $SiO_2$ (~110 nm). (c) Measured electrical resistance $R_{e,tot} = 2R_{e,C} + 2R_{e,\perp} + R_{e,\parallel}$ vs. distance $L$ between neighboring contact pads for varying SL film etched depths, at room temperature. (d) Calculated total area-specific resistance (including the Pt/SL contact resistance), $\rho_{etch}$ as a function of the etched depth. Each data point is averaged from the contact resistivities of five M-TLM devices with the same etch depth. The errors due to the fabrication uncertainties are calculated using a Monte Carlo method.[33]

# Supporting Information

# Uncovering Thermal and Electrical Properties of Sb₂Te₃/GeTe Superlattice Films

Heungdong Kwon[1], Asir Intisar Khan[2], Christopher Perez[1], Mehdi Asheghi[1], Eric Pop[2,3], Kenneth E. Goodson[1,3,*]

[1]Dept. of Mechanical Eng., Stanford University, Stanford, CA 94305, U.S.A.

[2]Dept. of Electrical Eng., Stanford University, Stanford, CA 94305, U.S.A.

[3]Dept. of Materials Science & Eng., Stanford University, Stanford, CA 94305, U.S.A.

[*]*E-mail*: goodson@stanford.edu

## 1. Time Domain Thermoreflectance (TDTR)

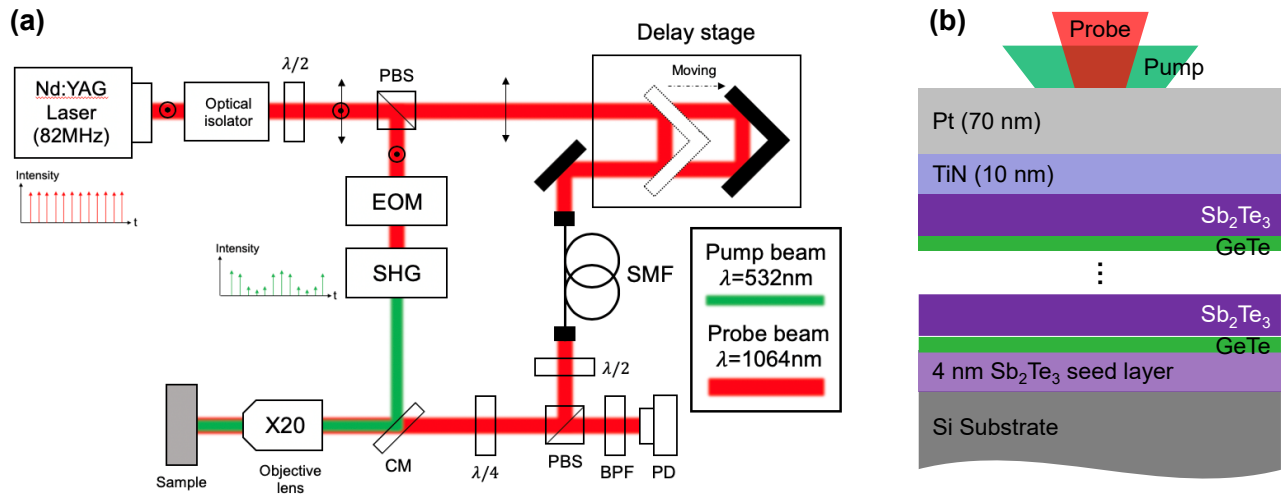

**Figure S1.** (a) Schematic of TDTR experiment set-up. (b) Schematic of the film stack used for TDTR measurements of the superlattice type phase change material (SL-PCM). The sample consists of (bottom to top): silicon substrate, 4 nm Sb₂Te₃ seed layer, repeating Sb₂Te₃/GeTe layers, 10 nm TiN capping layer, and 70 nm thick platinum transducer for TDTR measurement.

As depicted in Figure S1, our TDTR set-up use Nd:YAG pulsed laser emitting pulse train with 82 MHz repetition rate. The laser output that passes through an optical isolator is divided into the pump beam and probe beam paths. The pump beam pulse train is modulated with a sinusoidal envelope from 1 MHz to 10 MHz using an electro-optic modulator (EOM) and the 1064 nm wavelength is halved in the second harmonic generator (SHG). The modulated pump laser reflected from the cold mirror (CM) is projected onto the sample to induce modulated surface heating of the platinum transducer layer. The probe beam is time-delayed behind the pump pulse train by using the precise delay stage. The time-delayed probe beam goes through a single mode fiber (SMF) and is reflected on the polarized beam splitter (PBS). The beam is then circularly polarized through the quarter wave plate (λ/4). The reflected probe laser is detected by a high-speed photodetector (PD). The optical signal is measured by the lock-in amplifier wired to the PD. We monitor the ratio between in- and out-of-phase voltage signals from the lock-in amplifier as a function of the time delay between the pump and probe beams. This ratio signal is fitted with a multilayer heat diffusion model[1] to extract the unknown parameters.

## 2. Sensitivity Analysis

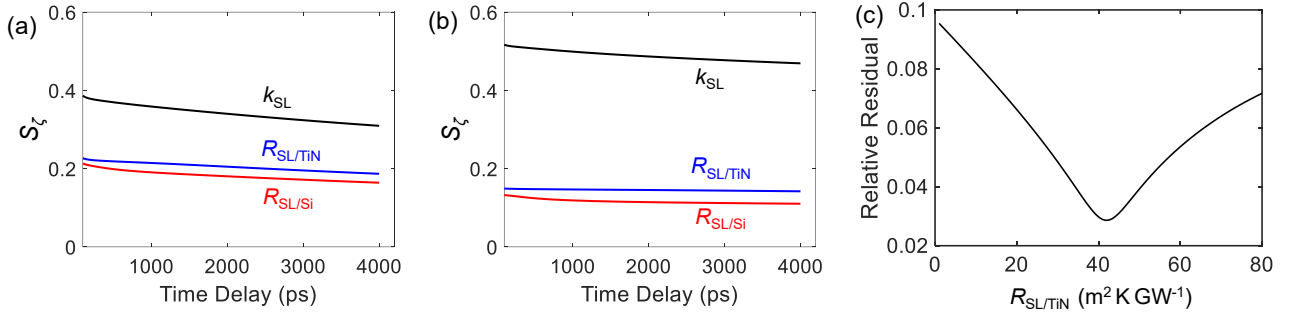

**Figure S2**. Measurement sensitivity of the TDTR signal to the thermal conductivity of SL ($k_{SL}$), to the thermal boundary resistance between TiN capping and SL ($R_{SL/TiN}$), and to the thermal boundary resistance between SL and silicon substrate ($R_{SL/Si}$). The sensitivity analysis is performed on (a) 14 nm and (b) 29 nm thick 4/1 nm/nm Sb$_2$Te$_3$/GeTe SL samples. The 4 MHz heating modulation frequency is used as the frequency in the experiment. (c) For a 29 nm thick sample, the relative residual of the fitting is plotted as a function of the assumed value of $R_{SL/TiN}$.

The sensitivity parameter $S_\zeta$ represents the relative change of TDTR output signal (voltage ratio) to the fractional change in a property of interest ($\zeta$). As shown in Figure S2a above, for the thinnest sample (14 nm thick SL), $R_{SL/TiN}$ and $R_{SL/Si}$ have a similar sensitivity coefficient because the heat wave penetrates all the way into the silicon substrate. In other words, they are equally weighted, which means that the thermal conductivity of the SL film bounded by these two interfaces is similarly weighted. Therefore, the TDTR measurement on this sample yields $R_{total}=R_{SL/TiN} + d_{SL}/k_{SL} + R_{SL/Si} \approx 86.1$ m$^2$ K GW$^{-1}$, where $d_{SL}$ is the total thickness of the SL. The measured intrinsic thermal conductivity of the SL $k_{SL} \approx 0.41$ W m$^{-1}$ K$^{-1}$ from the slope of the line fitting the total thermal resistances of four samples is used to reduce one unknown parameter in $R_{total}$ (main text Figure 1d). Using the intermediately thick SL (29 nm), of which $R_{SL/TiN}$, $k_{SL}$, and $R_{SL/Si}$ are sensitive to TDTR signals as seen in Figure S2b, we fit $R_{SL/TiN}$ to minimize the fitting error with the constraint of $R_{total}$ as shown in Figure S2c. The minimum error occurs at $R_{SL/TiN}$ = 41.7 ± 10.8 m$^2$ K GW$^{-1}$, confirming $R_{SL/TiN}$ is nearly four times larger than $R_{SL/Si}$, which is ~11.9 m$^2$ K GW$^{-1}$.

| Properties | Materials | | | |
|---|---|---|---|---|
| | **TiN** | **fcc-Ge₂Sb₂Te₅** | **Sb₂Te₃** | **GeTe** |
| $\rho$ **[g cm⁻³]** | 5.40 | 6.30 | 6.51 | 6.18 |
| $C_v$ **[MJ m⁻³ K⁻¹]** | 3.14 (Ref. 7) | 1.34 (Ref. 8) | 1.29 (Ref. 9) | 1.53 (Ref. 10) |
| $v_{s,L}$ **[m/s]** | 9638 (Ref. 11) | 3190 (Ref. 12) | 2757 (c-axis)[Ref. 13] | 2670 (c-axis)[Ref. 14] |
| $k_{bulk,lattice}$ **[W m⁻¹ K⁻¹]** | - | - | 0.8 (c-axis)[Ref. 15] | 2 (c-axis)[Ref. 15] |

**Table S1**. Summary of the reported values for the density, volumetric specific heat, and longitudinal sound velocity for TiN, fcc-Ge₂Sb₂Te₅, crystalline Sb₂Te₃, and crystalline GeTe. For Sb₂Te₃ and GeTe, the cross-plane (c-axis) longitudinal sound velocity and thermal conductivity data are referred to estimate the phonon mean free path in SLs.

## 3. Theoretical Calculation of Thermal Boundary Resistance (TBR)

The result shows ~47 % larger thermal boundary resistance than the reported data[2] between TiN and fcc-Ge₂Sb₂Te₅ (GST), $R_{GST/TiN} \approx 28$ m² K GW⁻¹. Several studies have indicated that the cross-plane acoustic velocity of the SL film could be suppressed significantly due to the creation of minibands and flattened phonon dispersion spectra in SLs.[3,4] To explain the larger $R_{SL/TiN}$ compared to $R_{GST/TiN}$, we calculate the theoretical TBR using the Diffuse Mismatch Model (DMM) with a gray approximation.[5]

$$R = \frac{C_{v,1}}{4\pi}\left(\frac{v_2 C_{v,2}}{v_1 C_{v,1} + v_2 C_{v,2}}\right)^{-1}$$

where $C_v$ is the volumetric specific heat, $v$ is the sound velocity, and the subscripts indicate the material on either side of the interface. The material properties used in the TBR calculation are presented in Table S1. The volumetric specific heat in Table S1 is calculated from the density in the table and the molecular specific heat given by the corresponding references. For the sound velocity in the model, we use the values of longitudinal sound velocity. Although this simplified model cannot predict an accurate TBR compared to the DMM using a full phonon dispersion, the approximation can provide insight into the trend of TBRs. For the cross-plane sound velocity of the SL film, we use the following equation given in Ref 6.

$$v_{SL} = \frac{d}{\sqrt{\frac{d_1^2}{v_1^2} + \frac{d_2^2}{v_2^2} + \left(\frac{Z_1}{Z_2} + \frac{Z_1}{Z_2}\right)\frac{d_1 d_2}{v_1 v_2}}}$$

where $Z = \rho v$ is the acoustic impedance, $d$ is the period of the SL, and the subscripts indicate the materials on either side of the interface (i.e. Sb₂Te₃ and GeTe). The estimated effective velocity across the SL is used as the parameter to calculate the thermal boundary resistance between TiN and SL-PCM with the DMM. As expected, the averaged volumetric specific heat (1.32 MJ m⁻³ K⁻¹) and longitudinal acoustic velocity (2739 m/s) of 4/1 nm/nm Sb₂Te₃/GeTe are not noticeably different from those of fcc-Ge₂Sb₂Te₅ (shown in Table S1), owing to their similar molecular composition. It was observed that the measured longitudinal sound velocity across a Sb₂Te₃/Bi₂Te₃ SL is ~10% reduced from its effective sound velocity.[6] In this study, we use the same scaling factor to estimate the decreased cross-plane longitudinal sound velocity of the 4/1 nm/nm $Sb_2Te_3$/GeTe film from the effective velocity. The ratio between $R_{SL/TiN}$ and $R_{GST/TiN}$ is then calculated to be ~1.27. Though the ratio of TBRs calculated using the simple DMM model underestimates the experimental values of TBR between SL/TiN relative to the reported data of TBR of fcc-GST/TiN film, we attribute the higher TBR of SL-PCM to the decreased sound velocity in SL film.

## 4. Minimum Thermal Conductivity

The minimum thermal conductivity model proposed by Cahill *et al*.[16] describes the thermal conductivity of a highly disordered crystal with a random walk motion of vibrational modes, as:

$$k_{min} = \frac{1}{2}\left(\frac{\pi}{6}\right)^{\frac{1}{3}} k_B n_a^{\frac{2}{3}} (v_l + 2v_t)$$

where $k_B$ is the Boltzmann constant, $n_a$ is the atomic number density, $v_l$ is the longitudinal speed of sound, and $v_t$ is the transverse speed of sound. Using this model, the theoretical lower bound of thermal conductivity for superlattices can be estimated.[17] To estimate the minimum thermal conductivity of SL explored in this study, first, we calculate the individual minimum thermal conductivities of $Sb_2Te_3$ and GeTe respectively, and then the minimum thermal conductivity of SL is estimated with their thickness ratio in the SL period unit. The atomic density is derived using the mass density and molar mass of each material and the sound velocity data of $Sb_2Te_3$ and GeTe are taken from Ref. 13 and Ref. 14, respectively.

## 5. Simkin and Mahan model (SMM)

To analyze our experimental data of the cross-plane thermal conductivity of superlattice (SL) films, we employ the Simkin and Mahan model (SMM),[18] which describes thermal transport in the SL films with SL period smaller than the coherence length of phonon. This approach reduces the SL system into a periodic one-dimensional mass-spring chain system where molecular mass of a molecule (e.g. $Sb_2Te_3$ or GeTe) that comprises the SL is considered. From the calculated band structure determined from the characteristic matrix, we estimate the thermal conductivities of SLs. The input parameters for SMM are the bulk phonon mean free path and the molecular mass ratio between the two materials in the SL. We consider the 4:1 thickness ratio of $Sb_2Te_3$ and GeTe in one period of SL for our case. However, this model assumes elastic scattering processes at the interface, which cannot account for imperfections or disorder at the interfaces present in our SL films, some of which can be seen in Figure 1b of the main text. Therefore, we modify the SMM as described in Ref. 19 to incorporate the impact of these imperfections. In addition, using the conventional kinetic theory, the estimated phonon mean free path across the SL films is ~1 nm (as explained in the main text), which is shorter than the period of the SL. In this case, SMM cannot predict the experimentally observed thermal conductivity in such SL. Therefore, to obtain a better estimate of phonon mean free path across our SLs, we fit the experimental thermal conductivity data using a modified SMM (M-SMM). The bulk phonon mean free path across the SL ($\lambda$), and the thermal boundary resistances *per interface* due to imperfections acting alone ($R_{imp}$) are used as the two unknown free parameters, see Figure 2b in the main text.

To obtain further insight, we plot (Figure S3) various thermal resistances of Sb$_2$Te$_3$/GeTe SL films (total thickness 60 nm) with varying period thicknesses for a fixed mean free path of 9.4 nm and four different values of $R_{imp}$ = 0, 1, 2, and 3 m$^2$ K GW$^{-1}$ (TBR per interface due to imperfections, acting alone). As the TBR due to the imperfection at the interface increases, the total thermal resistance calculated using the modified SMM ($R_{M\text{-}SMM}$) in the SL becomes larger than the thermal resistance estimated by SMM. For $R_{imp}$ ~3 m$^2$ GW$^{-1}$ (Figure S3d), a minimum thermal conductivity of SL (a maximum thermal resistance) cannot be observed, because of the dominance of the thermal resistance due to the imperfections in the overall resistance of SL film. This indicates that the phonon coherence phenomena may not be observed in SL films with lower quality interfaces (i.e., higher $R_{imp}$). At the same time, this further points to the good quality interfaces in our studied SLs.

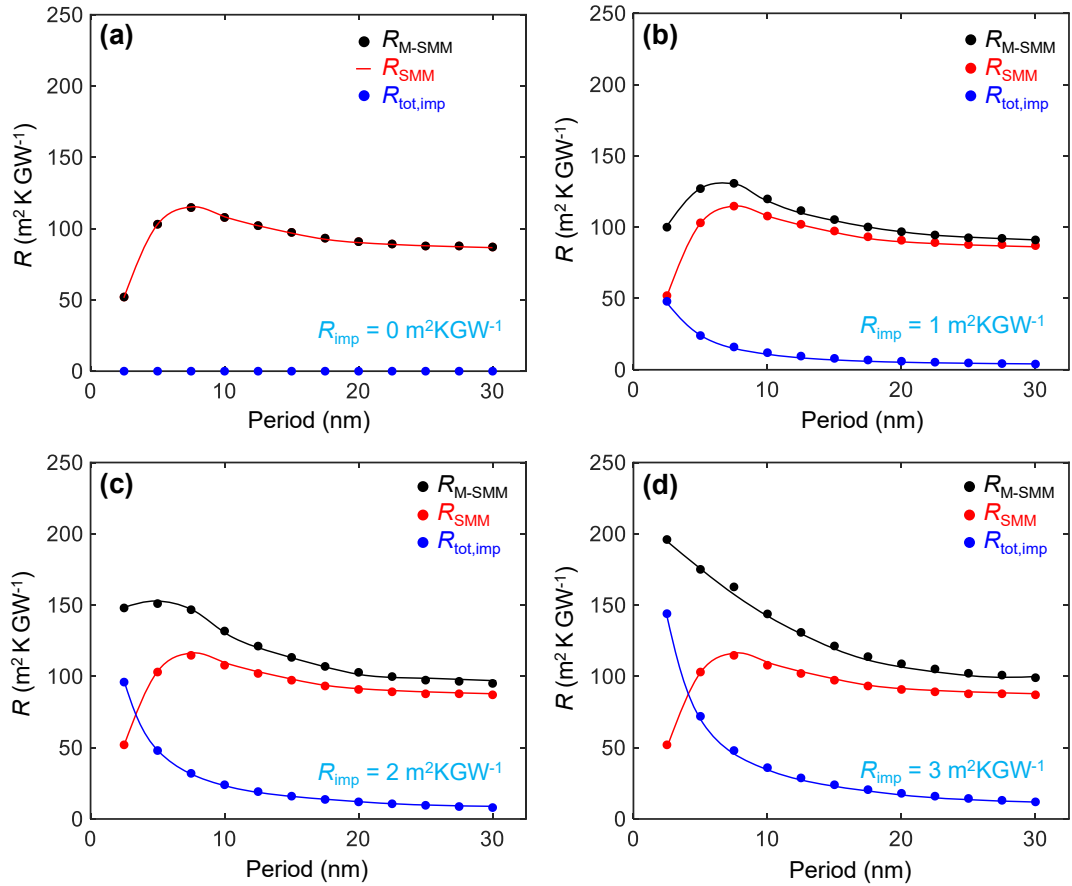

**Figure S3.** The calculated total thermal resistance of Sb$_2$Te$_3$/GeTe SL using modified SMM $R_{M\text{-}SMM}$ (black), the thermal resistance of SL estimated by SMM $R_{SMM}$ (red), and the total thermal resistance due to imperfections at the interfaces in the SL $R_{tot,imp}$ (blue) as a function of the period thicknesses for varying $R_{imp}$ = (a) 0 m$^2$ K GW$^{-1}$, (b) 1 m$^2$ K GW$^{-1}$, (c) 2 m$^2$ K GW$^{-1}$, and (d) 3 m$^2$ K GW$^{-1}$. Here, $R_{imp}$ is the thermal boundary resistance per interface in SL due to imperfections acting alone. $R_{\text{M-SMM}} = R_{\text{SMM}} + R_{\text{tot,imp}}$ where, $R_{\text{tot,imp}} = n R_{\text{imp}}$ and $n$ is the total number of the interfaces in SL.

## 6. Top view images of modified transmission line method devices and atomic force microscopy results

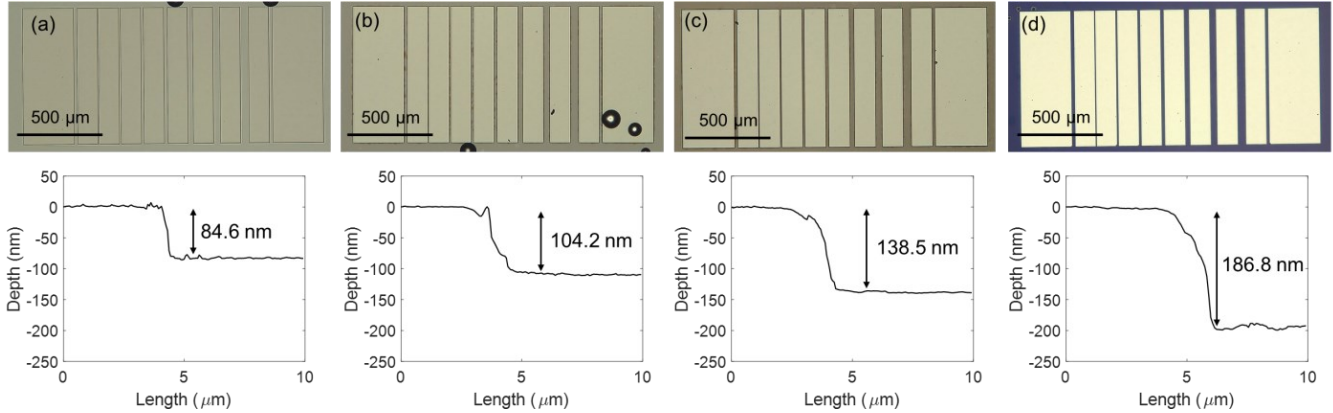

**Figure S4**. Top view images of modified transmission line method (M-TLM) devices with different etched depth and corresponding depth profiling results using atomic force microscopy, which measures the depth from top surface of platinum metal pattern to SL-PCM surface. (a) The device without SL-PCM etched, where the depth corresponds to Pt thickness, (b) with SL-PCM etched for 30 seconds, (c) with SL-PCM etched for 40 seconds, and (d) with SL-PCM fully etched, where the ~110 nm thick thermally grown $SiO_2$ is exposed (see cross-section in main text Figure 3b).

| Material | Electrical resistivity | | References |
|---|---|---|---|
| | In-plane ($\Omega\,cm$) | Cross-plane ($\Omega\,cm$) | |
| $Sb_2Te_3$/GeTe SL | $5.8 \times 10^{-4}$ | 1.1 | This work |
| $Sb_2Te_3$/GeTe SL | $4.8 \times 10^{-3}$ | 19.4 | Ref 20. |
| $In_2Se_3$ nanowire SL | $2 \times 10^{-3}$ | 40 | Ref 21. |
| $Sb_2Te_3$ | $1 \times 10^{-3}$ | $4 \times 10^{-3}$ | Ref 22. |
| $Sb_2Te_3$/GeTe SL | $7.8 \times 10^{-4}$ | - | Ref 23. |
| $Sb_2Te_3$/GeTe SL | $8.2 \times 10^{-4}$ | - | Ref 24. |
| $Sb_2Te_3$/$Ge_2Sb_2Te_5$ SL | $9.2 \times 10^{-4}$ | - | Ref 24. |
| $Sb_2Te_3$/$Bi_2Te_3$ SL | $7.7 \times 10^{-4}$ | - | Ref 25. |

**Table S2**. Reported data of electrical resistivities in the in-plane and cross-plane directions of different materials including the $Sb_2Te_3$/GeTe superlattice (SL) explored in this work. All materials are (poly)crystalline, although deposition conditions can vary between samples and between studies.

## Supporting References